# Implementing a neural network interatomic model with performance portability for emerging exascale architectures

**Saaketh Desai[1], Samuel Temple Reeve[2], James F. Belak[2]**

1. School of Materials Engineering and Birck Nanotechnology Center, Purdue University, West Lafayette, IN, USA 47906
2. Materials Science Division, Lawrence Livermore National Laboratory, Livermore, California, USA 94550

## Abstract

The two main thrusts of computational science are more accurate predictions and faster calculations; to this end, the zeitgeist in molecular dynamics (MD) simulations is pursuing machine learned and data driven interatomic models, e.g. neural network potentials, and novel hardware architectures, e.g. GPUs. Current implementations of neural network potentials are orders of magnitude slower than traditional interatomic models and while looming exascale computing offers the ability to run large, accurate simulations with these models, achieving portable performance for MD with new and varied exascale hardware requires rethinking traditional algorithms, using novel data structures, and library solutions. We re-implement a neural network interatomic model in CabanaMD, an MD proxy application, built on libraries developed for performance portability. Our implementation shows significantly improved on-node scaling in this complex kernel as compared to a current LAMMPS implementation, across both strong and weak scaling. Our single-source solution results in improved performance in many cases, with thread-scalability enabling simulations up to 21 million atoms on a single CPU node and 2 million atoms on a single GPU. We also explore parallelism and data layout choices (using flexible data structures called AoSoAs) and their effect on performance, seeing up to ~25% and ~10% improvements in performance on a GPU simply by choosing the right level of parallelism and data layout, respectively.

## 1. Introduction

Molecular dynamics (MD) simulations have been used for many years [1] to study metals and alloys, polymers, 2D materials, proteins, and more. Advances in computing power have allowed MD simulations of billions of atoms in the last decade [2], elucidating complex processes such as solidification [3,4] and plasticity [5], while also realizing microsecond timescales [6]. Such advances have been possible largely due to the advent of GPUs, necessitating the re-implementation of MD algorithms to extract maximal performance on both GPUs and increasingly hierarchical multi-core CPUs. These re-implementations have resulted in many GPU-accelerated MD codes [6–10], as well as calls for performance portability measures [11]. With exascale computing on the horizon [12], modeling and simulations tools can be increasingly used to solve ‘grand challenges’ in material science, such as understanding high-temperature superconductivity using quantum mechanical (QM) simulations, simulating energetic materials under extreme conditions at an atomistic level, and enabling additive manufacturing (AM) with multi-scale modeling of the processes involved.

Solving these ‘grand challenges’ requires efficient leverage of exascale computing resources via exascale-ready simulation tools. The Exascale Computing Project (ECP) is focused on achieving this by delivering an exascale computing ecosystem via co-design between hardware architectures, software stack development, and application development [13,14]. This ensures first that application requirements are met by hardware and software innovations and also that applications can effectively use cutting-edge hardware and software stacks. An important arm of the ECP initiative is within co-design centers across all three areas based on computational “motifs”: operations on particles, linear algebra, structured/unstructured grids, graph operations, etc. These co-design centers have two main outputs to facilitate co-design and interoperability: libraries and proxy applications. Software libraries can be used directly within production applications, while proxy apps are instead designed as a testbed for the main features of a complex physics application, aiming to understand and optimize computation, memory access, network usage, etc., separate from the full application code.

The Co-design center for Particle Applications (CoPA) (https://github.com/ECP-copa) addresses motifs within all particle-based simulations, across many application areas, ranging from atoms to galaxies, via the development of software libraries and packages targeted towards accelerating the key motifs for each application. CoPA is developing the PROGRESS and BML libraries for linear algebra in quantum MD for chemistry and materials science [15], as well as the Cabana library for particle operations [16] within n-body gravity for cosmology, particle-in-cell methods for plasma physics, and classical MD for materials science, the focus of this work. Cabana was designed to leverage on-node parallelism from the Kokkos library [17] for cross-platform performance, adding particle specific algorithms and MPI communication for particles.

CabanaMD is a classical MD proxy app that uses the Cabana and Kokkos libraries and has shown performance portability across various hardware for a Lennard-Jones (LJ) interatomic model. At nearly 100 years old, the LJ interatomic model is still widely used in qualitative studies and for van der Waals interactions in quantitative predictions, the simplicity of the kernel also proving valuable for assessing one extreme of MD performance. For most materials however, more complex interatomic models are required. Machine learned, data driven, and neural network models aim to offer near-quantum level accuracy in predicting energies and forces for a wide range of systems, while retaining the linear scaling offered by classical MD. Neural network interatomic models [18,19] in particular offer unique and complex computational kernels, enabling many avenues for improvement of the computational cost, as well as expansion of features within both CabanaMD and Cabana. These neural network potentials (NNP) have notably recently gained traction by accurately simulating phase transitions in disparate materials [20–23]. Neural network models rely on the flexible functional form of a neural network, with thousands of tunable parameters, to approximate complex potential energy surfaces. In this work, we re-implement an NNP in CabanaMD, demonstrating use of the Kokkos and Cabana libraries to achieve performance portability, and scalable performance of a neural network interatomic model, both across CPUs and GPUs.

The remainder of the paper is as follows. We first describe neural network and related, relatively new interatomic models in Section 2. In Section 3 we examine the NNP kernels in detail, briefly describe the CabanaMD proxy application and its use of the Kokkos programming model and the Cabana particle toolkit to enable co-design for MD codes, as well as discussing the main ways in which the code was modified from a current LAMMPS implementation. On-node strong and weak scaling with the CabanaMD proxy application is demonstrated on both multi-core CPU and GPU hardware for various materials, comparing our code with the previous CPU-only implementation in Section 4. In addition, we explore parallelism and data layout improvements using unique Kokkos and Cabana library features and run simulations of up to 21 million (M) atoms on a single CPU node and up to 2M atoms on a single GPU on state of the art pre-exascale architectures. Finally, we conclude and discuss impact and continuing work in Section 5. Our work highlights a path forward for data-driven and machine-learned interatomic models in classical MD at the exascale.

## 2. Neural network interatomic models

Machine learned and data-driven interatomic models offer the opportunity to perform nearly-quantum accurate MD simulations while retaining the *scaling* of a standard interatomic model. These next-generation models are a departure from traditional, empirically motivated, parametric models, e.g. embedded atom method  [24], where the number of parameters and functions are fixed. Non-parametric interatomic models instead use descriptors, which characterize the atomic environment of each atom, together with a regression model. These non-parametric descriptors are more mathematically-motivated, although still physically derived and hand-tuned in implementation,  and include symmetry functions [18], bispectrum coefficients [25], , and others [26–28]. Notably, all versions of non-parametric interatomic models are systematically improvable by adding additional basis functions or more nodes in the neural network [27], unlike standard, fixed-parameter models. The physics of the system, including any translational, rotational, and permutational invariances are included in the descriptors. These descriptors can also be learned using machine learning techniques such as an additional neural network or complex convolutional/residual networks to learn the environment of each atom, as opposed to hand-tuning a set of descriptors, thus requiring even less human intervention [29–31]. Assessing and improving the performance of learned-descriptor-based models is fundamentally different from descriptor-based models and will be subject to a future study.

Across non-parametric models, the mapping technique that relates the structural descriptors (learned or prescribed) to the observed quantities (e.g. energies, forces, and stresses from a QM calculation) varies widely. For instance, SNAP uses a linear regression mapping of bispectrum coefficient descriptors [25], while symmetry function descriptor based (Behler-style) NNP uses a simple feedforward neural network as the mapping [18]. For the learned-descriptor based NNPs, additional neural network(s) are used as mappings [29–31]. We note that these models are still under active development, with many iterations for each descriptor, mapping, and combination. Having chosen a descriptor and a mapping, the interatomic model parameters are typically obtained through a least-squares fitting procedure, converging parameters to a local optimum. Improvements on this strategy include combining local optimization techniques with global optimization for hyperparameters and using techniques such as genetic algorithms [32].

While not the focus of this work, it is important to note that the accuracy and range of environments represented by non-parametric models depends strongly on the QM data used to train the model, as with their parametric counterparts. The training data set must be large and diverse enough such that the model is primarily interpolating while in use. Traditional approaches use training data with range and sampling pre-defined by domain experts; however, recent advances show successful application of principal component analysis (PCA), active learning, and transfer learning techniques to ensure reasonable sampling of configurations across the training set, and that each configuration provides added value, which may not be obvious to the expert [33–35].

In this work, we focus on the performance of an NNP that uses symmetry functions as descriptors, which are Gaussian functions of neighbor distances with radial or angular character, with each symmetry function varying in the parameters of the Gaussian and the elements involved. The flexibility and accuracy of neural network models is achieved at a high computational cost relative to standard interatomic models, with comparative performance for many non-parametric models recently benchmarked [36]. Recent work has suggested improvements to NNP descriptors, aiming to reduce the number of symmetry functions required to describe multi-element neighborhoods [37], and thus improve performance. We aim to instead improve performance by implementing a cross-platform, thread-scalable version of a neural network model in the CabanaMD proxy app.

# 3. Model Implementation

## 3.1 Neural network interatomic kernels

To understand the scope for improvements in the Behler-style NNP, we look at the computational kernel in detail. Figure 1(a) shows a traditional Lennard-Jones (LJ) kernel that shows the loop over each atom *i* and each of its neighbors *j* to compute the force on each atom. Note that the energy compute step is shown only for comparison and is not strictly required to compute the trajectory for an LJ kernel (and most often computed only occasionally). Figure 1(b) contrasts the simple LJ kernel with the computationally complex NNP kernel, broken down into 3 steps. Step 1 consists of a loop over each atom *i* and each of its neighbors *j*, but also has an added loop over each descriptor *k* to compute all radial atomic environment descriptors. Not shown is a similar angular computation with one extra loop over neighbors of neighbors. Step 2 is unique to the NNP kernel and involves matrix multiplications that forward-propagate the computed symmetry functions (descriptors) through the neural network to calculate the atomic energy. Thus, the energy calculation is necessary for the dynamics of the system, unlike a standard model. Step 3 is again similar to the LJ kernel in that it consists of a loop over atom *i* and each of its neighbors *j* to compute forces, but again involves an extra loop over descriptor *k*, and contains both the gradient of the neural network (w.r.t descriptors) and the gradient of symmetry functions (w.r.t coordinate). Note that while the neural network is not particularly large or deep (~30 x 2), each atom has its own neural network, resulting in many small sets of matrix multiplications (see Figure 1(c)). The use of machine learning libraries such as Tensorflow/Pytorch could be explored to gain additional speedup, but is left as future work, both because it is a minor portion of the overall computation, as seen later, and because these libraries are focused primarily on large, deep networks.

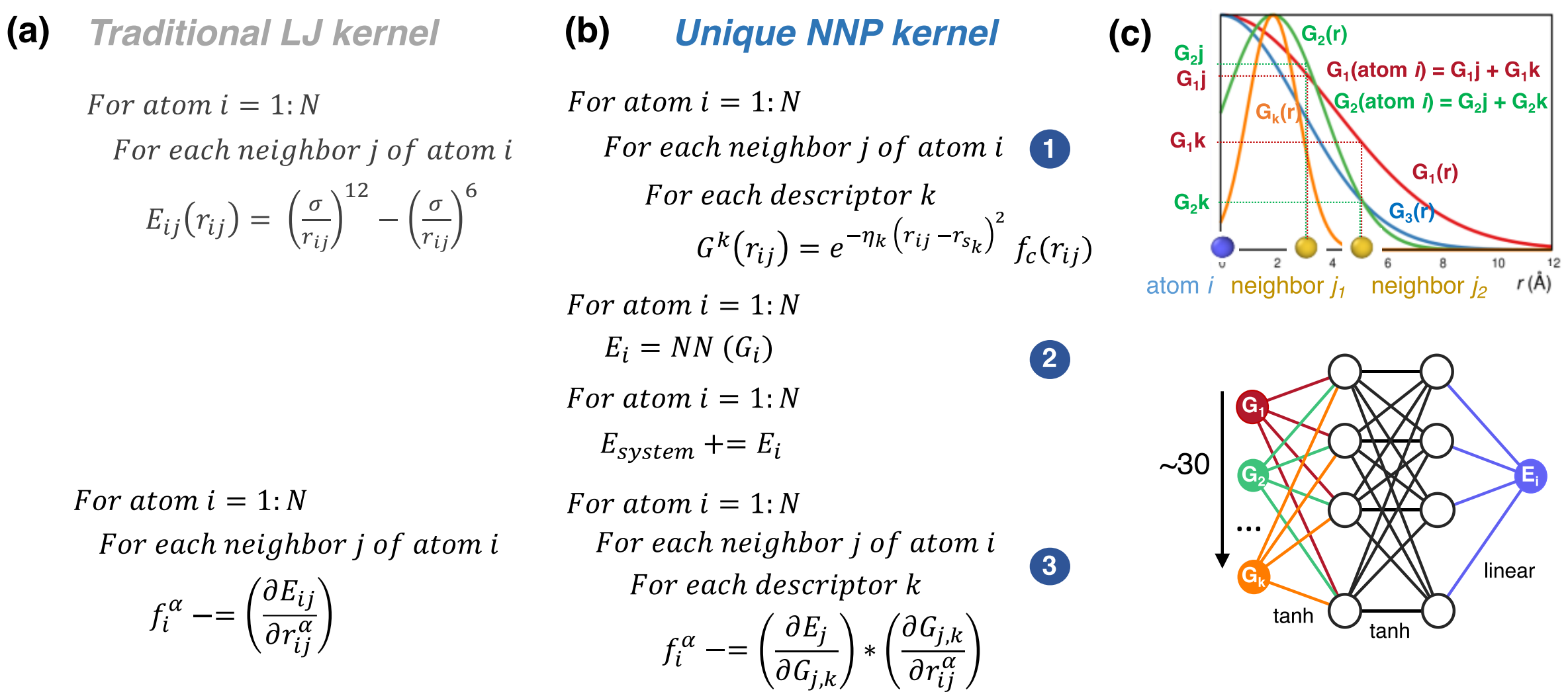

*Figure 1: (a) Traditional Lennard-Jones model showing the main loops over atoms and neighbors to compute energy and forces, (b) neural network interatomic model broken down into 3 steps: descriptors, neural network, and force calculations, and (c) representative NNP descriptors and schematic atomic neural network used to predict energy and forces.*

## 3.2 CabanaMD proxy application

CabanaMD (https://github.com/ECP-copa/CabanaMD), our testbed for re-implementing the neural network interatomic model, contains representative units from production MD codes, with flexibility for parallelism and data layout improvements. CabanaMD inherits the modular design and Kokkos implementations of the ExaMiniMD proxy app (https://github.com/ECP-copa/ExaMiniMD) [17]. CabanaMD primarily uses the Cabana library for flexibility of particle algorithms and data layout, with underlying on-node performance portability from Kokkos. The complete software stack is shown in Figure 2(a), highlighting that Cabana is a domain specific, direct extension of Kokkos. For parallelism across hardware in CabanaMD, the **Kokkos::parallel_for** is used for threading across atoms, except within loops involving atom neighbors, where the **Cabana::neighbor_parallel_for** provides loops over both atoms and neighbors (where either can be serial or threaded), built on the hierarchical parallelism of Kokkos. Portable memory for particle data uses array-of-structs-of-arrays (**Cabana::AoSoA**), built directly on **Kokkos::Views** with an additional compile-time array dimension that can map to a SIMD/SIMT instruction length we refer to as the vector length, see Figure 2(b). The AoSoA is intermediate between array-of-structs (AoS) and structs-of-arrays (SoA), more common data layouts which require a complete tradeoff between data locality and sequential data access, and thus provides both layout flexibility and benefits of both AoS and SoA for intermediate vector lengths. **Cabana::AoSoAs** are used for particle positions, velocities, forces, and types, additionally storing symmetry functions and gradients of atomic energy with respect to symmetry functions for the NNP kernel. **Kokkos::Views** are also used for type-based storage, for potential parameters, masses, symmetry function parameters, and neural network parameters (weights and biases). This functionality has enabled CabanaMD to demonstrate portable performance for an LJ model across Intel Xeon, IBM POWER9, AMD Epyc, and ARM ThunderX2 CPU compute nodes, as well as NVIDIA P100/V100 GPUs. As shown in Section 4, the use of appropriate levels of parallelism and data layouts for given hardware increases performance by up to 25% for the NNP kernel.

### 3.3 Modifications from previous implementation

In this work, we re-implement the NNP from the n2p2 package (https://github.com/CompPhysVienna/n2p2), a library-based LAMMPS [38] implementation with a model for water provided [19,39]. We also test our implementation using the NNP models developed for Ni, Cu, Si, Ge, Li, and Mo as a part of a recent comparative benchmarking study [36,40]. Several key changes were necessary in translating the n2p2 implementation to CabanaMD, particularly for GPU architecture: (i) move from large storage classes and structs to minimal **Kokkos::Views** and **Cabana::AoSoAs**, (ii) replace explicit OpenMP pragmas with **Kokkos::parallel_for** and **Cabana::neighbor_parallel_for**, (iii) rewrite some classes of object-oriented code to use less memory, and (iv) replace code that is not GPU-compatible, such as swapping out vector iterators with loops over **Kokkos::Views** (v) Split computationally expensive kernels and use Kokkos and Cabana constructs to consider various levels of parallelism (parallelism over only atoms vs atoms and neighbors).

The n2p2 LAMMPS implementation was created and optimized extremely well for standard CPU computing, avoiding floating point operations by precomputing and storing variables wherever possible. In addition, a focus on MPI-parallelization, with OpenMP included mainly for MPI+X, indicates that the code was not intended for significant threading performance, meaning that a direct re-implementation of n2p2 in CabanaMD would result in extremely low utilization of GPU resources. To avoid this, an almost complete reversal was necessary: store and move as little data as possible, recomputing where necessary. The n2p2 implementation neatly exposed potential points of parallelization and we used this parallelization within Kokkos and Cabana parallel constructs with thread-safe atomics as needed for symmetry function and force updates. However, the data layout needed to be fully inverted, from structs of all atomic and neighbor properties to minimal AoS or AoSoA in Cabana. It was further necessary to replace some of the object-oriented code structure with kernel-based parallel code. Finally, we note that CabanaMD depends on the n2p2 library for initialization and other functionality that is not performance critical.

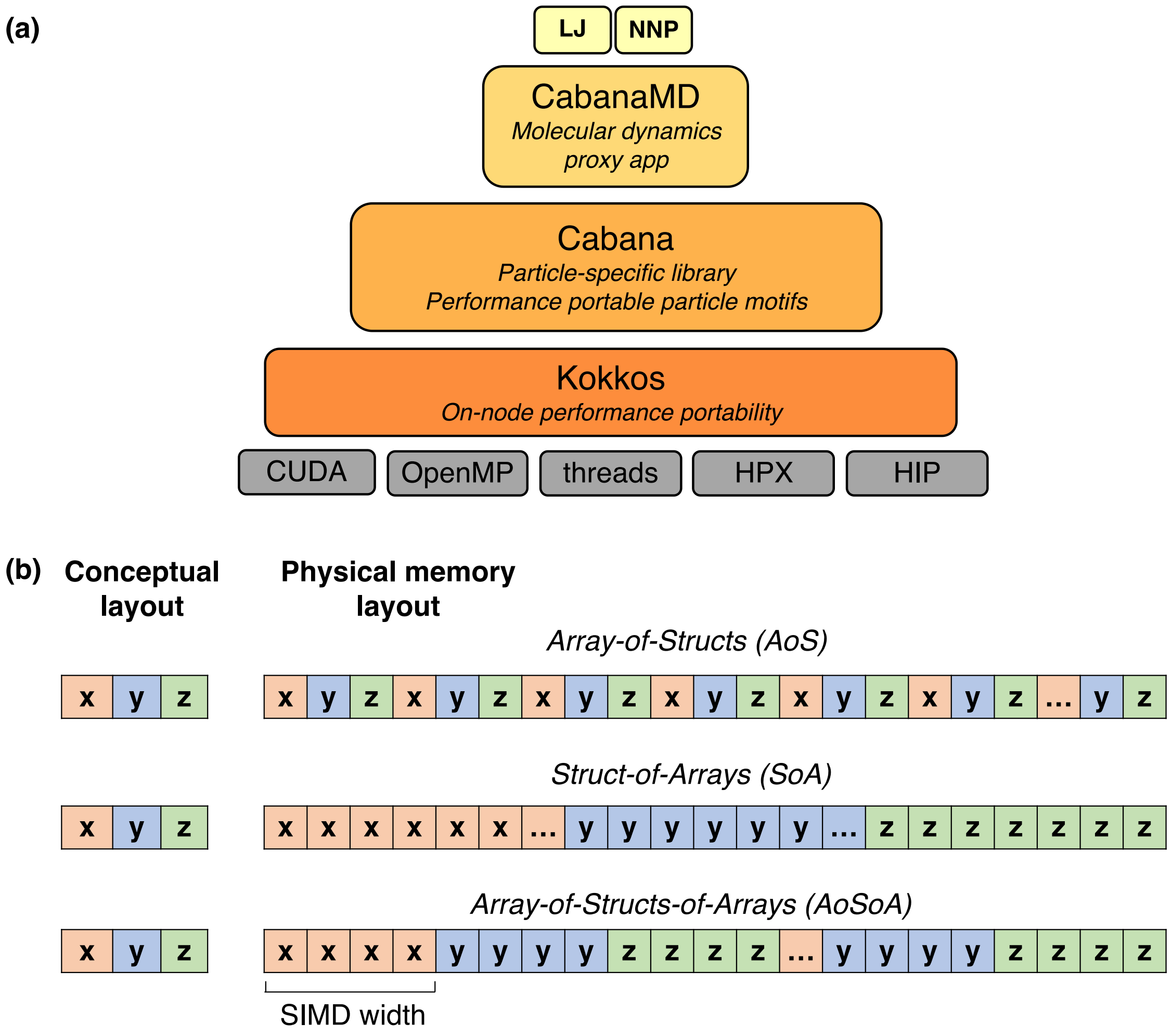

*Figure 2: (a) Software stack representing CabanaMD-NNP (b) Cabana AoSoA data structure allowing flexibility in data layout, compared to standard AoS and SoA.*

# 4. Performance

We show on-node performance with both many-core CPU and GPU architectures on a current leadership-class supercomputer, the pre-exascale Lassen machine at LLNL, see Figure 3. CPU scaling behavior is documented for an IBM POWER9 node (with 44 cores per node and 4 threads per core), while GPU results are shown for a single NVIDIA V100 GPU. This architecture matches the Summit and Sierra supercomputers at Oak Ridge National Lab and LLNL, respectively. We note that the code used for this work is directly available within a branch of the Github project: https://github.com/ECP-copa/CabanaMD/tree/NNP_OnNode. The NNP functionality will be continually updated through the master branch.

## 4.1 CPU OpenMP performance

Figure 3(a) shows strong scaling performance with increasing number of OpenMP threads on a single POWER9 node, comparing the n2p2 LAMMPS implementation with CabanaMD for a nickel test system [40]. The n2p2 implementation shows saturating performance due to a memory-bound implementation with large structs to store atom and neighbor properties, including symmetry functions and derivatives. In contrast, our thread-bound implementation shows increasing performance with number of threads, showing linear scaling (with some reduction in that scaling, yet still linear for hyperthreading). This results in a greater parallel efficiency (~70%) for our implementation and allows systems of up to 21M atoms to be run on a single node. The n2p2 algorithms show better performance for the smallest numbers of threads, where the heavy compute and low resource nature of the task result in performance gains from a storage-based approach, surpassing exposed parallelism. However, even in using hybrid MPI+OpenMP parallelization (a large focus of their effort), the n2p2 implementation is more limited in system size by memory than CabanaMD. Ultimately, CabanaMD shows a speedup of ~3x for large systems with many threads while n2p2 is faster by ~2x when restricted to a few threads, see Fig. 3(a). Weak scaling from Figure 3(b) reinforces these points, where n2p2's superior performance for smaller systems with a single thread is overtaken by the parallelism from Cabana and Kokkos in CabanaMD. Figure 3(c) and 3(d) similarly compare n2p2 and CabanaMD strong and weak scaling for $H_2O$ [39], and while we see that n2p2 is faster than our implementation, CabanaMD shows better thread scalability. We attribute this to the larger cutoff for the $H_2O$ potential (6.4Å vs 3.9Å for Ni), significantly increasing the number of neighbors for which computations are required. Thus, the $H_2O$ model represents a cross-over point, where the memory and thread-based approaches are similar, for the POWER9 hardware. We note that in other cases, particularly older CPUs with slower and/or smaller memory, CabanaMD can be considerably faster, even for $H_2O$. Supplementary Figure S1 shows cases for Intel Xeon (Broadwell), where the CabanaMD speedups are ~12x for Ni and ~3x for H2O, using the maximum number of threads.

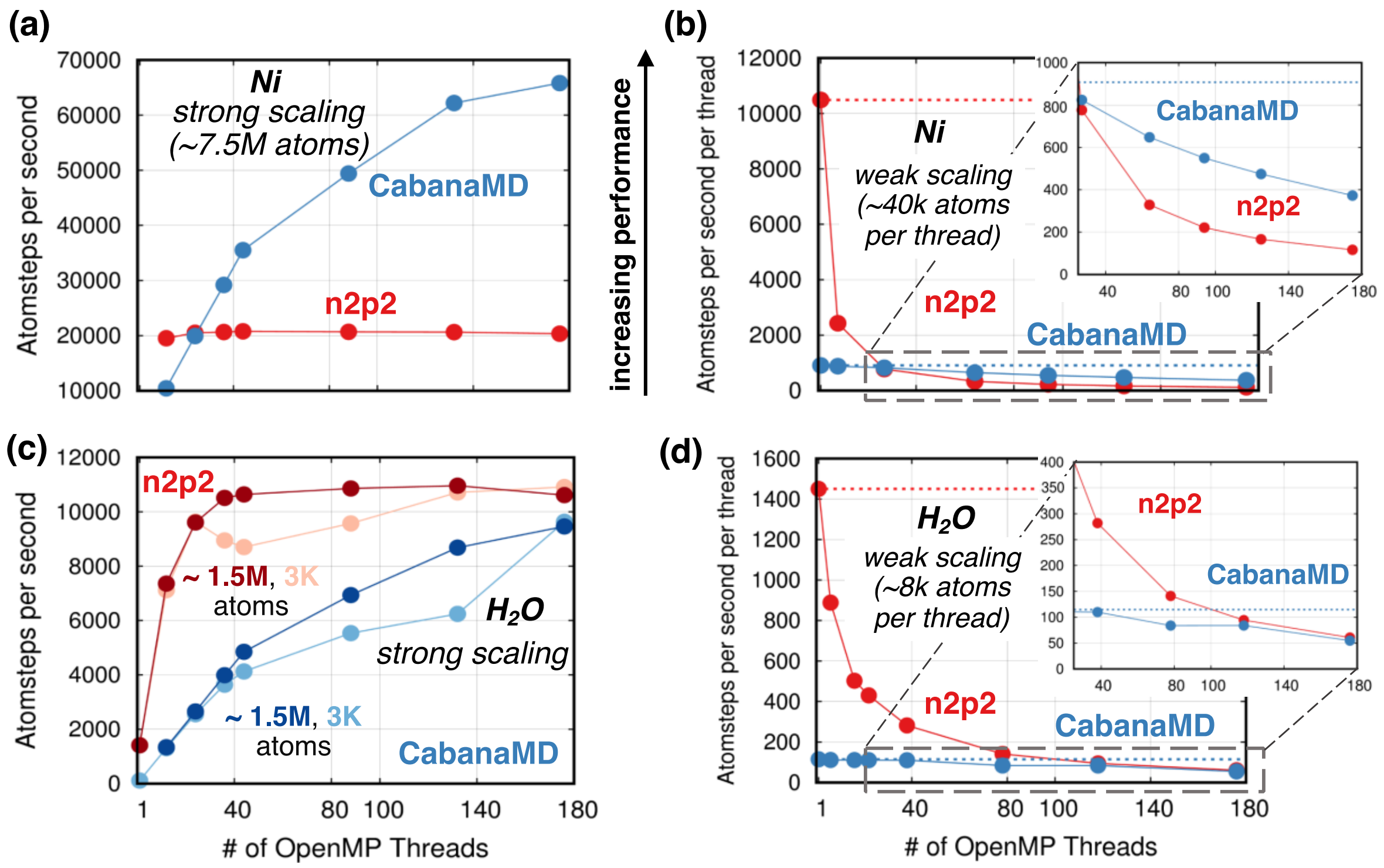

*Figure 3: (a) Strong and (b) weak scaling for Ni contrasting the n2p2 and CabanaMD implementations, evaluated on a single IBM POWER9 node. (c) Strong and (d) weak scaling for $H_2O$ further comparing codes.*

## 4.2 GPU performance

The use of the Kokkos and Cabana libraries provides us a single-source solution for performance portability. Figure 4 highlights the performance of our implementation on the GPU, where we observe a speedup of ~4x for $H_2O$ compared to the CPU, with this speedup being approximately an order of magnitude for Ni (and ~12x , ~11x and ~9x for other single element systems such as Cu, Mo and Ge). Energy conservation for all systems is documented in Supplementary Figure S2.

We again stress that we are able to run up to 21M atoms on a single CPU node and 2M atoms on a single GPU, again by avoiding a memory-based approach, which would severely limit the system sizes and speeds achievable on the GPU. Crucially, implementation decisions detailed in Section 3.3 resulted in not only faster GPU simulations, but faster large-scale CPU simulations as well.

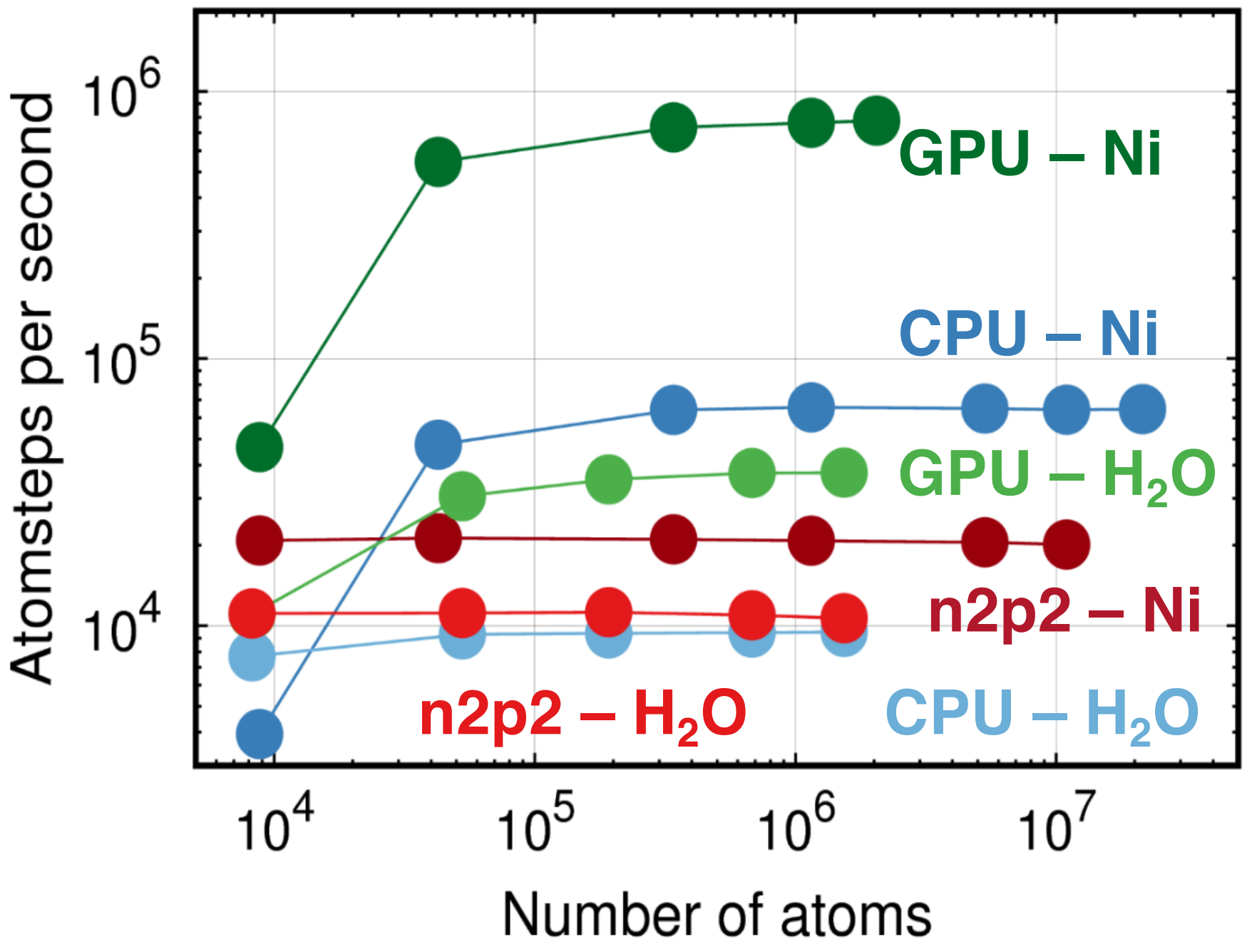

*Figure 4: Performance comparisons for Ni and $H_2O$, with CabanaMD on a single NVIDIA V100 GPU and CabanaMD with 176 threads on an IBM POWER9 CPU node, and the n2p2 implementation with the same CPU.*

### 4.3 Parallelism and data layout improvements

The CabanaMD implementation shown above largely retains the algorithms used by n2p2 in the current LAMMPS implementation. However, the power of the Cabana implementation is in enabling easy exploration of new algorithms and parallelization strategies via separation of concerns, as well as exploration of data layouts via flexible AoSoAs. To better understand the performance of our implementation, we first breakdown the kernel timings into the three steps from Fig. 1, shown in Figure 5. We observe that the neural network is a minor part of the computation, similar across hardware, with the calculation of forces taking up to 60% of the NNP time per MD step. This highlights that while the NNP kernel is novel and includes a new compute step (propagation through the neural network), optimizations targeted at other models, up to and including the LJ kernel, are still most useful for improvements.

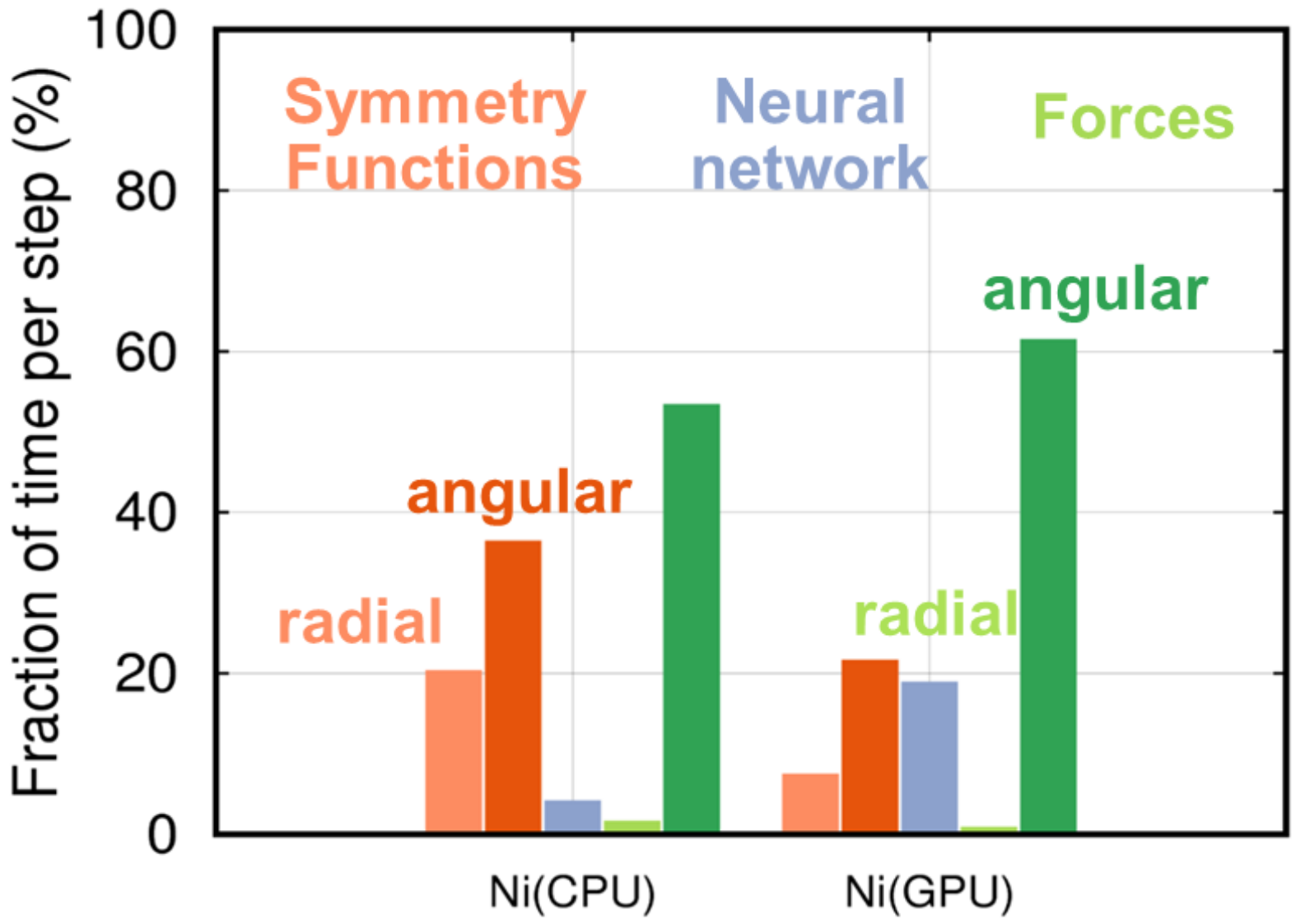

*Figure 5: Fraction of time taken by each portion of the CabanaMD-NNP compute kernel for a ~1.5M atom Ni system in computing symmetry functions, propagating those symmetry functions through the atomic neural networks, and computing forces. Note that the force contribution from the neural network is computed and counted within the force bar.*

The main algorithmic changes involved recomputing rather than reading stored values throughout the code. In addition, we split the computation of symmetry functions into separate radial and angular symmetry function kernels and exposed greater parallelism, using **Cabana::neighbor_parallel_for** to parallelize computations over both atoms and neighbors. This functionality directly uses hierarchical parallelism, mapping loops over atoms and neighbors to the multiple levels of compute and memory hardware, which we compare and contrast with flat parallelism (threaded parallelism over atoms only). Use of the hierarchical parallelism require an additional atomic update of the symmetry functions, in addition to atomic update of the final forces that is always necessary. Figure 6 shows that the use of hierarchical parallelism results in ~8% performance improvement on the CPU for large systems, while significantly reducing performance for a small system with ~1000 atoms. At the smallest sizes there is not enough total work to make the additional overheads of multiple levels of parallelism worthwhile; however, for most systems on the CPU the higher exposed parallelism outweighs the necessary additional atomic operations, primarily because the total number of threads is small. On the GPU, the use of hierarchical parallelism significantly degrades performance, where in this case the much larger number of threads results in significant contention between them, making it difficult to leverage the greater exposed parallelism. There is again a cross-over for the smallest systems on the GPU where, without the considerable thread contention, neighbor parallelism improves performance. Throughout sections 4.1-4.2, we thus use hierarchical parallelism on the CPU and flat parallelism on the GPU using a simple command line flag, a capability possible due to use of flexible Kokkos and Cabana constructs.

A third level of parallelism, again using **Cabana::neighbor_parallel_for**, parallelizing computations over atoms, neighbor, and angular neighbors is also available, but resulted in only a ~1% performance improvement for a 1000 atom system and slightly reduced performance for larger systems, as compared to parallelizing over neighbors. This third level of parallelism is mapped to vector units directly, not amenable to random access neighbor operations which dominate these simulations. However, for other materials, system sizes, interatomic models, or algorithms this feature could be relevant.

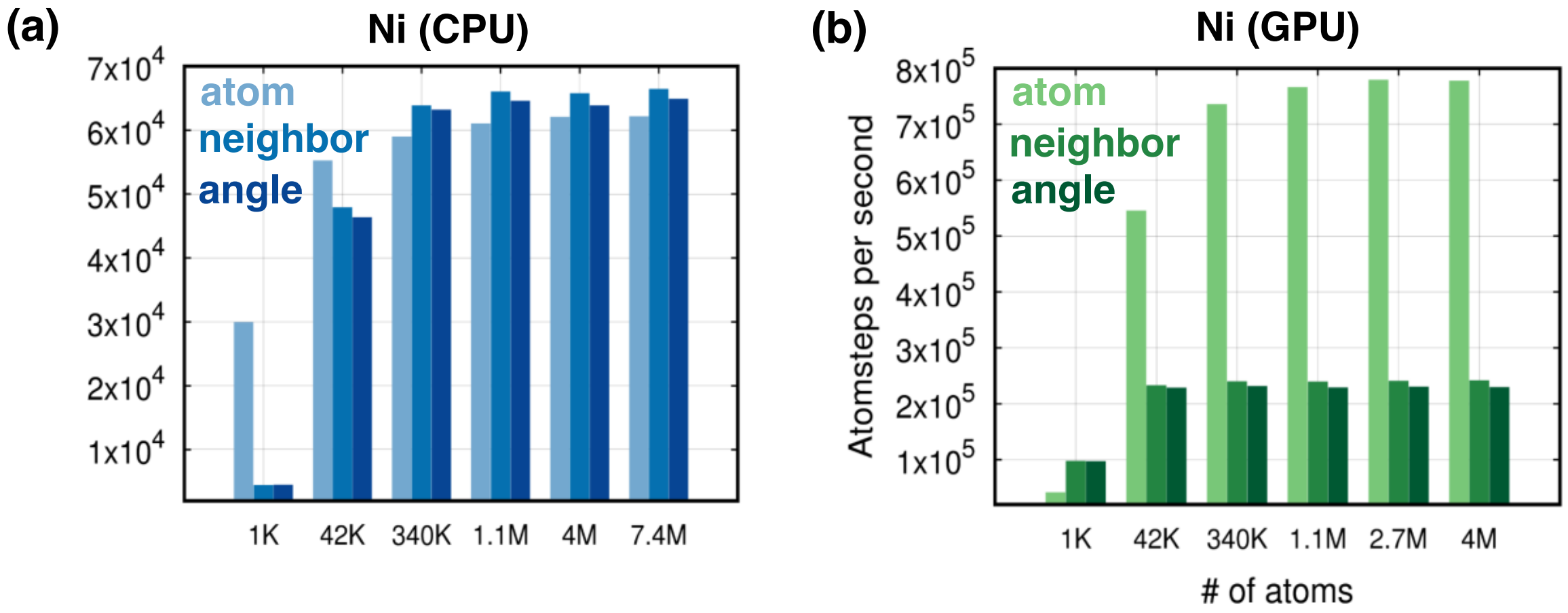

*Figure 6: Performance for Ni compared across various levels of parallelism for a ~1.5M atom Ni system on (a) a single IBM POWER9 node and (b) a single NVIDIA V100 GPU.*

We can also improve performance by choosing the appropriate data layout for the architecture. Figure 7 shows performance gains obtained (on CPU and GPU) by varying the vector length for the **Cabana::AoSoA**, in addition to demonstrating the performance tradeoffs in using a single combined AoSoA vs using multiple separate AoSoAs (for all particle properties, including NNP-specific arrays), varying two levels of tradeoffs between data locality and consecutive access. It should be noted that each **Kokkos::View** underlying the **Cabana::AoSoA** automatically uses the more performant layout for data access with a given architecture: row major (**Kokkos::LayoutRight**) for CPU and column major (**Kokkos::LayoutLeft**) for GPU.

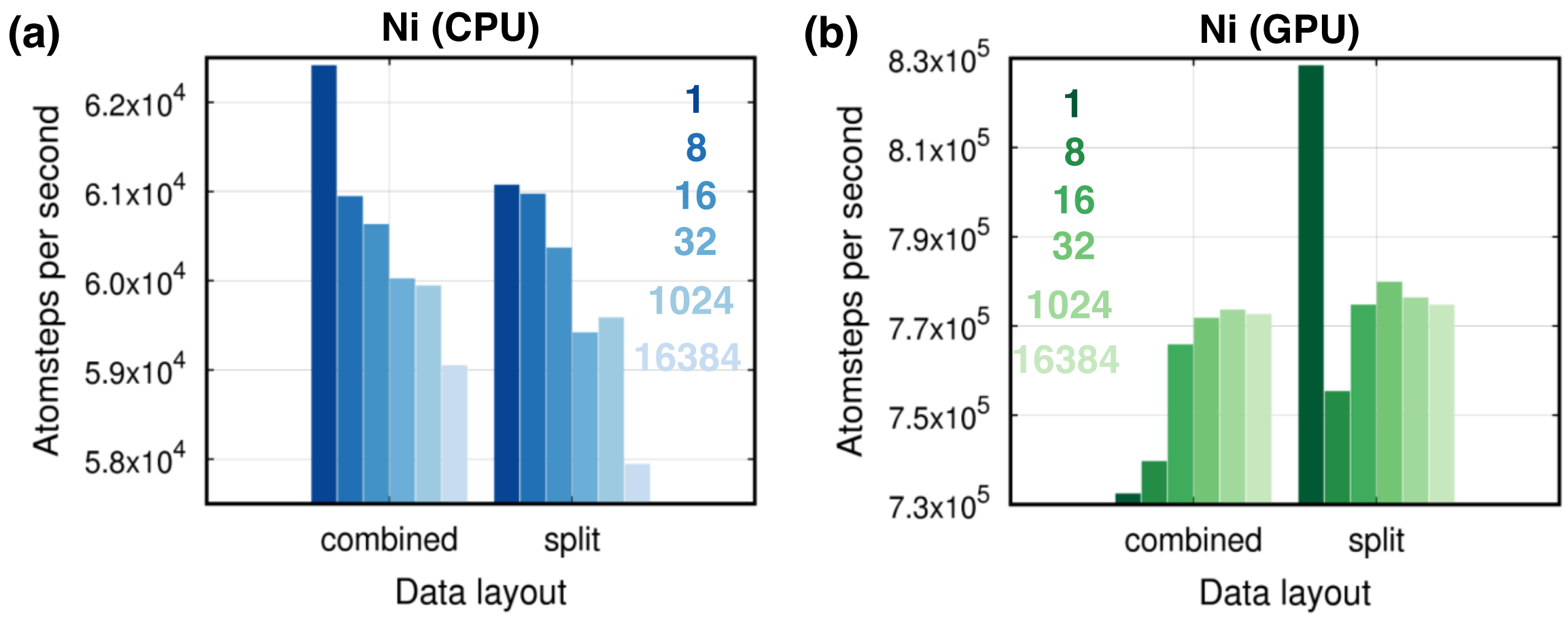

*Figure 7: Performance for Ni compared across various vector length and AoSoA sizes on (a) a single IBM POWER9 node and (b) a single NVIDIA V100 GPU for a ~1.5M atom Ni system*

Overall, the trends are reversed between CPU and GPU. For the CPU, a vector length of 1 (AoS) gives best performance and degrades as the vector length increases, as data locality is reduced and without significant vectorization, Figure 7(a). Because the majority of the simulation time is from random access (neighbor list) force computations, the vectorizable benefits from the intermediate vector length AoSoA layout are not currently taken advantage of; however, it is possible to rewrite MD kernels to be partially vectorizable [41,42]. Breaking the AoSoA into individual particle properties does not significantly change performance but is consistently slightly slower. The combined AoSoA with vector length of 1 results in ~4% better performance over the slowest data layout on the CPU and we use this layout throughout Section 4. On the GPU, performance increases with increasing vector length, other than a spike for an AoS with split particle properties, Figure 7(b). This highlights the competing effects on the GPU; the dominant angular force loop does not vectorize, while the loops over symmetry functions and neural networks can to some degree. For the overall trends, the increasing vector lengths improve the data parallel non-neighbor loops, without significant slowdown from moderate reduction in data locality. The split AoS case is, however, able to speed up the angular force loop with much better data locality (only half of the particle data is required for this kernel), outweighing the slowdown of the vectorizable loops due to non-consecutive layout. We see up to a 10% improvement between the fastest and slowest GPU data layout. We used the combined AoSoA with a vector length of 32 throughout in order to use the same codebase for all results (with the exception of Figure 7), as the split AoSoA required some code changes. We finally note that vector lengths of 16 and 32, close to the warp size, give nominally the same performance as vector lengths approaching an SoA.

## 5. Conclusions

We have demonstrated significant performance improvements for large-scale MD with a neural network interatomic model within the CabanaMD proxy app, as compared to the existing LAMMPS-based n2p2 library. CabanaMD-NNP is built on n2p2, for many NNP utilities, and the Cabana library for particle methods, itself built on the Kokkos library for on-node performance portability, enabling continuing exploration of parallelization and data layout/access on emerging computing architectures. These improvements are relevant to the n2p2 package [19], as well as the myriad of Behler-style NNP implementations [43–49] and other new models using NN in different ways [50,51]. Further, learned-descriptor-based non-parametric models, containing completely unique kernels, will pose challenges to improving performance not seen here and will be subject to future work [29–31].

The main ideas of this work apply to any current CPU-only code looking to utilize GPU resources: with a performance portable library or programming model, data storage and movement must be avoided in favor of re-computing. In our case, this manifests itself in our choice to re-compute symmetry functions and derivatives, as opposed to storing these large data structures for each atom and neighbor. In addition, exposing as much parallelism as possible and determining which levels to thread over is integral. Importantly, these changes not only enable GPU performance, but also improve CPU multi-threading in many cases, albeit at the expense of performance for some small systems. CabanaMD-NNP is available at https://github.com/ECP-copa/CabanaMD/ and the version used for these results is archived within the same repository (https://github.com/ECP-copa/CabanaMD/tree/NNP_OnNode).

The ability to simulate millions or billions of atoms with nearly-quantum accurate models, using performant MD implementations such as ours, will significantly expand the set of scientific problems within the reach of classical MD. This work highlights the ability to do so, with multi-node scaling a focus moving forward. We intend to use the CabanaMD-NNP implementation to investigate complex materials without currently available models: solid-solid phase transitions, AM-rate solidification processes, physical and chemical phenomena in energetic materials, and microstructural evolution in f-block metallic alloys. In addition, for NNP models that have already been developed, CabanaMD-NNP enables unprecedented large-scale MD simulations of complex processes such as recrystallization in phase change memory applications [23] and diffusion in amorphous solid-state battery electrolytes [52].

## Acknowledgements

The authors acknowledge helpful discussion and support from Stuart Slattery. Work performed under the auspices of the U.S. DOE by LLNL under contracts DE-AC52-07NA27344 and supported by the Exascale Computing Project (17-SC-20-SC), a collaborative effort of the U.S. DOE Office of Science and the NNSA.